# Design of Liquid Impregnated Surface with Stable Lubricant layer in Mixed Water/Oil Environment for Low Hydrate Adhesion

Abhishek Mund, Amit K Nayse, Arindam Das[*]

School of Mechanical Sciences, Indian Institute of Technology (IIT) Goa, GEC Campus, Farmagudi, Ponda, Goa, 403401, India

*Email: arindam@iitgoa.ac.in

**Abstract:**

Clathrate hydrate is a naturally occurring ice-like solid which forms in water phase under suitable temperature and pressure conditions, in the presence of one or more hydrophobic molecules. It also forms inside the oil and gas pipes leading to higher pumping cost, flow blockage and even catastrophic accidents. Engineered surfaces with low hydrate adhesion can provide an effective solution to this problem. Liquid impregnated surfaces is one such example of engineered surfaces which has already shown tremendous potential in reducing the nucleation and adhesion of solids. Here we report the design and synthesis of liquid impregnated surfaces with extremely low hydrate adhesion under the mixed environment of oil and water. The most challenging aspect of designing these surfaces was to stabilize a lubricant layer simultaneously under the water and oil. A detailed methodology to make such lubricant stable surfaces from theoretical perspective was described and experimentally validated for lubricant stability. Experimental measurements on such surfaces showed extremely low hydrate accumulation and one order of magnitude or more reduction in hydrate adhesion force.

## 1. Introduction:

Clathrate hydrates are solid, crystalline solids in which hydrogen-bonded water (host) molecules form molecular cages with diverse configurations such as pentagonal and hexagonal faces, around low-molecular-weight species. To put it another way, gas molecules (guests) are trapped in water cavities (hosts) made up of hydrogen-bonded water molecules. Since its discovery in 1810 by Sir Humphry Davy, Clathrate hydrate particularly gas hydrate were extensively investigated[1]. Villard[2] was the first to discover that hydrates are a broad category of compounds that contain a variety of guest molecules such as methane, ethane, and propane, in 1888. Clathrate hydrates were found both in nature and man-made systems. In nature, clathrate hydrates form under high pressure and low temperature at the bottom of the sea, in that water is the host and methane is the main guest molecule[1]. In manmade systems, presence of this hydrates are significantly observed in oil and gas infrastructures, where it forms both inside and outside of fluid transport systems such as pipes, valves etc. Hammerschmidt discovered hydrate in industrial gas flow lines[3] in 1934, when particle build-up caused line obstruction. Hydrate forms at the water-hydrocarbon interface in typical oil/gas pipelines due to the favorable environmental[4] conditions in such system. Moderately high pressure of 5 MPa and comparatively low temperature below 10℃ present in such systems are perfectly suitable for hydrate formation. Hydrate phase once form, rapidly grows into to small particle, particle agglomerates, solid hydrate deposit and eventually leading to hydrate plug inside the pipelines. Formation of hydrate particles and hydrate plugs creates a major flow assurance problem and sometimes leads to severe accidents. They can choke or hinder flow lines, slowing operations and, in the worst-case scenario, collapsing pipelines and equipments[5]. The safe removal of clogged hydrates and the prevention of hydrate plugging are achieved by utilizing a substantial proportion of deep-water fluid flow resources. Exploring and interpreting the pipeline-plugging method of hydrates is significant and pragmatic in the context of economic risk management's goal of preventing hydrate blockage[6,7]. To mitigate the challenges originating from hydrate formation, anti-agglomerate, thermodynamic inhibitor[8], mechanical de-pressurization system[9] and heating systems are employed in the industries[1]. These techniques, meanwhile, are expensive, consume a significant amount of energy and they could possibly have adverse environmental impact. For example, methanol an inhibitor of hydrate used in large quantity for effective result. It not only increases the material cost (methanol) but also necessitate a complicated and involved separation process to get rid of these chemicals from oil and gas phases. Hence it is worth to find out some environment friendly, energy independent, passive approach to effectively address this problem. In the past decade, curiosity in passive anti-hydrate coatings that can prevent hydrate growth directly on pipe walls or reduce hydrate adhesion has surged. Two significant passive anti-hydrate coating categories[10], namely superhydrophobic and polymeric coatings—are frequently being investigated recently by the scientist working in this area.

For past several years, engineered surfaces such as Bio-inspired Superhydrophobic lotus leaf surfaces[11,12], superoleophobic[13], pitcher plant Slippery Liquid Infused Porous Surfaces (SLIPS)[14,15] and Lubricant Impregnated Surfaces (LIS)[16,17] were widely studied for their superior performance across different engineering applications. These surfaces have shown tremendous potential in reducing ice-adhesion[18,19], reducing corrosion[20,21], reducing drag forces[16], facilitating drop-wise condensation[22,23], increased heat transfer in condensation process[24] etc. Modified solid-liquid or liquid-liquid contact area and low inherent adhesion forces are prime reason behind such superior performance shown by these surfaces. Due to the low adhesive force and restricted contact area with liquid phases, solids condensing or nucleating from liquid phase have showed minimal adherence on such surfaces. These surfaces thus expected to provide similar performance in reducing hydrate adhesion with solid pipe-walls in a energy independent way. In recent years a no of research articles has shown superior performance of superhydrophobic surfaces and under oil superhydrophobic surfaces in reducing hydrate adhesion[25]. The functional durability of superhydrophobic surfaces' remains a major concern for their use in various applications. Presence of a stable air layer within the surface textures of Superhydrophobic surfaces (SHS), is critical for its various superior properties. However, due to it's high compressibility, low density and low viscosity, this air layer is lost under the droplet impact, water condensation, high pressure liquid environments, and thus rendering these surfaces inefficient or functionally damaged[26]. LIS and SLIPs surfaces on the other hand have their texture filled with an impregnated dense and viscous lubricating liquid layer. This layer efficiently eliminates some of these issues faced by SHS and expected to help in outperforming SHS. Liquid Impregnated Surfaces are textured surfaces that allow a lubricating fluid, such as a liquid, to remain in a thermodynamically stable state inside the roughness, under specific environments (liquid or gas). The presence of this lubricating liquid layer inside the texture of these surfaces over time is critical to their long-term functional durability, which is required for their usage in real-world applications. SLIPS on other hand are micro/nanoporous surfaces with a lubricant layer that repels other fluids that are immiscible with the lubricant and has a weaker affinity for the solid than the lubricant. The long-term and thermodynamic stability of lubricant layers for SLIPS surfaces under a range of environments was never a priority. Rather, a thick coating of lubricant was commonly injected on textured surfaces to achieve minimal adhesion between a liquid and a solid surface through creating a physical barrier. The SLIPS surface thus does not ensure the thermodynamic stability of lubricant layer. In absence of such stability lubricant oil can be replaced from the surface texture by the environment fluids, making those surfaces functionally damaged. Compared to the SLIPS surfaces, LIS expected to perform better in reducing solid adhesion due to the presence thermodynamically stable lubricant layer. Properly designed LIS have already shown to reduce the ice-adhesion[18,19], reduce corrosion[27] and reduce drag forces[16]. LIS even being one of the high performing engineered surfaces still not been investigated to reduce the hydrate adhesion. The actual use of the LIS in oil & Gas pipes, necessitate lubricant stability and hydrate repellency in two fluid environments at the same time. Only a few research in the field of LIS have looked at the thermodynamic stability of the lubricant layer in equilibrium with a drop of a probe liquid (to be repelled) in an air environment[28]. To the best of our knowledge stability analysis and fabrication of Omni phobic[29] LIS surface that repels both oil and water never being attempted. Hence it is worth evaluating these surfaces as potential solution to the hydrate problems through systematic theoretical analysis and experimental work.

In this report, we carried out theoretical analysis and constructed a regime map for designing LIS which will not only retain its lubricant layer in both water and oil phase but also reduce the hydrate adhesion force to exceptionally low values. The thermodynamic stability of a lubricant layer in an ambient liquid is studied for the first time. To measure the hydrate adhesion at atmospheric pressure, cyclopentane water hydrate system was selected as it mimics' the gas hydrate formation mechanism[30]. LIS required for such system requires omniphobic characteristics where lubricant oil must be stable inside the textures under both water and oil (cyclopentane) environment. A thorough theoretical framework and regime map based on the equilibrium condition, were constructed to anticipate the configuration of a multiphase interface consisting of textured solid and three immiscible fluids (Lubricating liquid, two environment fluids). The role of contact angle, surface chemistry and surface roughness parameter on the stability of the impregnated layer was theoretically investigated considering a static environment. Subsequently, theoretical predictions on lubricant stability were verified by experimental observations. To the best of our knowledge, Omni phobic LIS surfaces and lubricant layer stability analysis under submerged conditions have not been comprehensively addressed. Furthermore, no direct experimental evidence has been presented to support the equilibrium-based theoretical stability theory. This is the first investigation to provide such a full regime map for constructing Omni phobic LIS having a stable impregnated layer in both the water and oil phases, along the experimental validation. Analysis performed here can be well extended to other systems such as water in gas, gas in water system as well. Lithographically produced silicon micro-pillars (with and without nanospike) of different post spacings and nano spiky surfaces on silicon wafer were used in the current study to measure hydrate adhesion. Adhesion force measurements were performed to quantify the attractive interplay between the hydrate and LIS. Hydrate adhesion measurements carried out on the designed stable LIS surfaces with micro, nano and micro-nano solid textures showed almost one, two and three order of magnitude reduction respectively compared to the smooth controlled surface.

## 2. Theoretical analysis: Designing Stable Liquid Impregnated Surfaces under two immiscible fluids environment

Previous works on LIS stability, very limited in number, were entirely focused on stability of lubricant under a specified probe liquid. Selected lubricants were immiscible with probe liquid and remained inside the texture of LIS under the environment of probe liquid. In case of hydrate phobic LIS surface, selection of lubricant, surface chemistry and the surface texture have to be such that lubricant (has to be immiscible with water and oil phase) has to be stable under two liquids (oil and water) simultaneously. The purpose of this theoretical analysis is to create the regime map which will give the conditions corresponds to various multiphase interfacial configurations involving three immiscible fluids (including lubricant). Ultimately, conditions suitable for low hydrate adhesion will be determined and it will act as a guideline to fabricate such LIS surfaces that will significantly lower solid-water contact and reduce hydrate adhesion. Textured surfaces with square micro posts and a rectangle array were considered for theoretical analysis. This type of geometry is easy to analyze and fabricate with an optical lithography-based microfabrication process. Consider a solid surface that has square posts affixed to it (see figure 1).

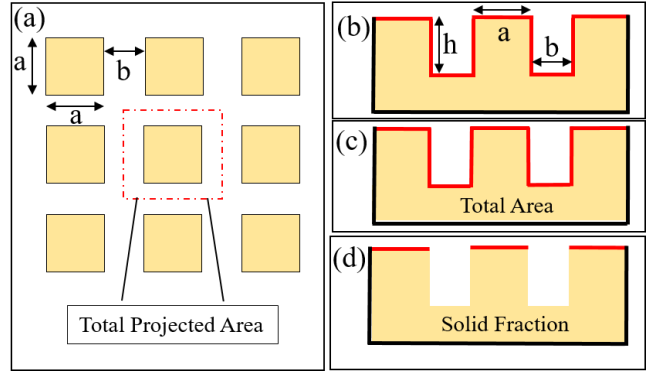

Figure 1. Schematic of representation of square post textured surface (a) Total projected area (b) dimensions (c) Total area (d) Solid fraction

Theoretical analysis is performed considering key geometrical properties such as $r_s$ (the ratio of total area to projected surface area), solid fraction $\varphi$ (the ratio of emerged surface area to projected surface area), width ($a$), edge-to-edge spacing ($b$), height ($h$), and critical contact angle ($\theta_c$). Equation 1, 2 and 3 gives the expressions of $r_s$, $\varphi$, and $\theta_c$ for micropost surfaces respectively, as shown earlier.

$$r_s = \left[\frac{(a+b)^2 + 4ah}{(a+b)^2}\right] \quad (1)$$

$$\varphi = \left[\frac{(a)^2}{(a+b)^2}\right] \quad (2)$$

$$Cos(\theta_c) = \left[\frac{1-\varphi}{r_s-\varphi}\right] \quad (3)$$

The principle of energy minimization was used to identify the thermodynamically stable interfacial configuration under different conditions defined by the relationship between surface roughness and surface chemistry parameters. The energy which was minimized here is the overall interfacial surface energy between distinct solid and liquid phases. This energy is the sum of the products of interfacial areas and particular specific interfacial energies of all interfaces between different phases. Geometry and surface chemistry are the sole determinants of these two terms. To construct the stability criteria and regime map for multiphase interfaces on LIS surfaces, appropriate representations of geometric characteristics and surface chemistry must be defined and compared. Specific surface energy values between liquid phases and solid surfaces are necessary to determine total surface energy. These specific surface energies (interfacial forces) are a direct articulation of intermolecular forces. These interfacial surface forces balance each other at three phase contact lines in an equilibrium state. This equilibrium state involving one smooth solid and two fluids is represented by equilibrium or young's contact angle given by Young's equation for smooth surfaces (shown in the figure 2). To properly describe the interfacial energy interactions involving probe liquid, environmental liquid and solid surface, an appropriate notation of this contact angle was used. As shown in the figure 4, the notation $\theta_{ws(cp)}$ denotes the contact angle of fluid phase *w* on a solid (s) surface in an environment of phase *cp*. Other contact angles, such as $\theta_{os(cp)}, \theta_{os(w)}$, are also calculated, involving the lubricating oil phase denoted by the letter ***o***.

The table below (see table 1) shows three possible configurations of the interface involving lubricant (o), environment/probe liquid (*cp*), and solid textured surface. Here $r_s$ is the total area roughness, $\varphi$ is a solid fraction normalized with the projected area. $E_{C1}$, $E_{C2}$, and $E_{C3}$ are total interfacial energy per unit area for C1, C2, and C3 configuration, respectively. At static conditions, this lubricant oil expected to be stable inside the texture if total surface energy associated with either of configurations given by figure b and c of table 1 is the lowest compared to other two configurations. Otherwise lubricant will not be stable and it will be replaced by environmental fluids. Total surface energies for all three configurations (shown in figure a to c of table 1) are function of surface energies and roughness properties.

Table 1 Schematic wetting configuration and their interfacial energy & criteria

| | Configurations | Total Interface energy per unit area | Equivalent Criteria |
|---|---|---|---|
| (a) | 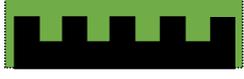 | $E_{C1} = r_s \gamma_{s(cp)}$ | $E_{C1} < E_{C2}, E_{C3}$ <br> $\theta os(cp) > \theta c$ |
| (b) | 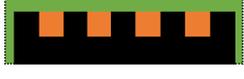 | $E_{C2} = (r_s - \varphi)\gamma_{os} + \varphi\gamma_{s(cp)}$ <br> $+(1-\varphi)\gamma_{o(cp)}$ | $E_{C2} < E_{C1}, E_{C3}$ <br> $0 < \theta os(cp) < \theta c$ |
| (c) | 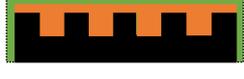 | $E_{C3} = \gamma_{o(cp)} + r_s\gamma_{os}$ | $E_{C3} < E_{C2}, E_{C1}$ <br> $\theta os(cp) = 0$ |

To find out the condition for which a particular interfacial configuration will be stable, energy associated with that configuration was set to be least compared to other two configurations. This leads to two inequalities. These inequalities gave the required stability conditions in terms of surface energy parameters and surface roughness parameters.

For example, if the state C2 has the lowest total interfacial energy, following inequalities were obtained,

$$E_{C2} < E_{C1} \Leftrightarrow \frac{(\gamma_{s(cp)} - \gamma_{os})}{\gamma_{o(cp)} > (1-\varphi)/(r_s - \varphi)} \quad (4)$$

$$E_{C2} < E_{C3} \Leftrightarrow \gamma_{s(cp)} - \gamma_{os} - \gamma_{o(cp)} < 0 \quad (5)$$

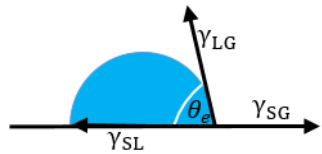

Figure 2. Liquid droplet on a homogeneous solid

From Young's equation of equilibrium contact angle on smooth surfaces (see figure 2), equations can be further simplified and expressed in terms of contact angles through following inequality,

$$0 < \theta os(cp) < \theta c \quad (6)$$

Here $\theta os(cp)$ represents contact angle of lubricant oil on smooth solid surface inside the environment of liquid denoted by $cp$. Similarly, conditions for which other two configurations will be thermodynamically stable, can be obtained and listed in table 1. Similar analysis can be performed to find out the conditions relevant to stability of lubricant oil layer inside the water environment. Theoretical analysis as described above, was performed for both $cp$ and $w$ environment fluid phases and results are combined to create a regime map (see figure 3). It depicts all possible configurations with associated conditions, expressed in terms of contact angles and the critical contact angle $\theta c$, a purely surface roughness parameter.

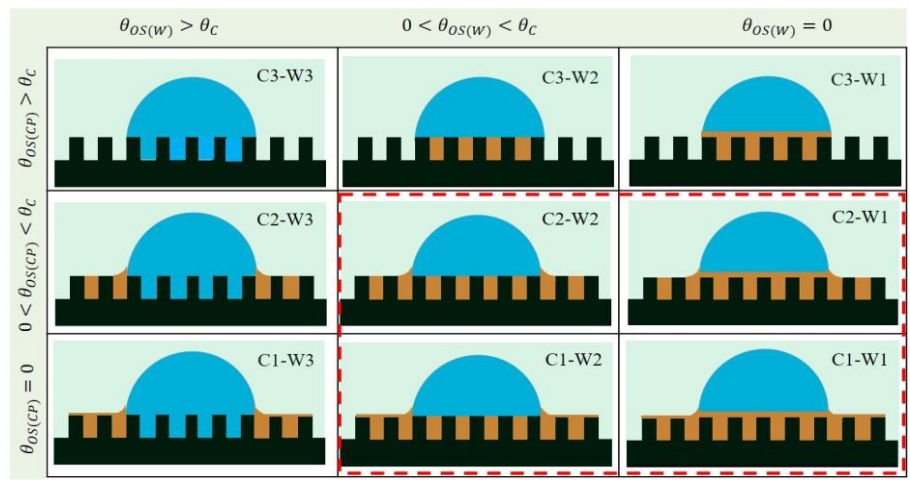

Figure 3 Possible Thermodynamically stable states of water drop placed on a LIS under environmental liquid.

Interfacial configurations indicated within red dotted lines are desired to have simultaneous stability under water and cyclopentane phase. Adhesion between the solid surface and solid phase (such as hydrate) originating from either of these fluids will depend on the contact area and adhesion between the solid and respective fluid phase. Among these four different states, configuration C1-W1 is the most desirable to achieve a LIS with an extreme repellency to both oil and water. The solid posts in this case are not in immediate contact with the probe liquid and environment liquid, thus expected to show substantial repellency with both the liquid phases and low adhesion with solid originating from either of these fluids. The solid posts in the other configurations, C1-W2, C2-W1, and C2-W2, are in direct exposure to probe liquid, environment liquid, and both liquids respectively, making them less desirable than C1-W1 when repellency to both fluids are concerned. Solid originating from dark blue phase and light blue phases expected to have low adhesion in configuration C2-W1 and C1-W2 configuration respectively and comparable to solid adhesion corresponds to C1-W1 configuration. Configuration C2-W2 is least desired as post tops are in contact with both fluids. However, the repellency and adhesion reduction in states with exposed post tops (in direct contact with liquids) can be significantly higher if the spacing between posts in those configurations are large enough. Theoretical analysis shows that the relationship between the interfacial interactions (between solid, lubricant, environment liquids) and the solid surface's geometry are key to design stable LIS surface with omniphobicity. These parameters have to be chosen very carefully to satisfy conditions as depicted in regime map (see figure 3) for specific set of environment fluids.

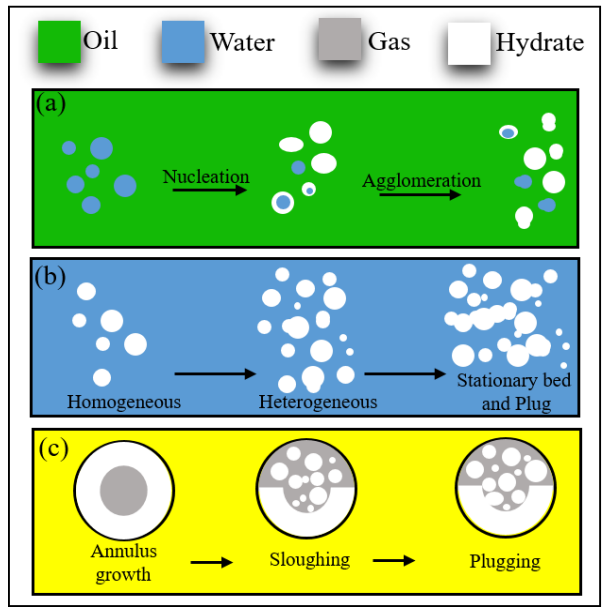

Figure 4 shows a graphic illustration of the different flow regimes that can occur inside oil and gas pipes. The regimes shown are (a) oil, (b) water, and (c) gas dominated systems

Hydrate adhesion on the inner surface of pipelines, comes under the topic of solid-solid adhesion. However, depending on the operating conditions, liquid-liquid, solid-liquid and solid-gas interaction or adhesion also became important in the hydrate adhesion. The environment has a strong impact on the process of hydrate adhesion growth. For an instance, in an environment where oil dominates and water is dispersed as drops, hydrate particle nucleates first on oil water interface, then agglomerates on pipeline surfaces (see figure 4). Homogeneous and heterogeneous hydrate particles develop stationary beds in the presence of water before clogging the pipe flow. The pipe becomes plugged when the hydrate particles begin to develop in the annular pipe and fill the internal space, in a gas dominated environment. These fundamental interactions will result in varying quantities of adhesive forces between hydrate and pipe walls. Thus, the properties of the medium can greatly alter the solid-solid adhesion. Current theoretical analysis performed on water-cyclopentane system are general in nature and it can be extended to all above mentioned configurations with selection of appropriate parameter corresponds to such systems

## 3. Results and Discussion:

**3.1 Fabrication of LIS:** Design and fabrication of LIS were carried out based upon the theoretical calculations described in the previous section. Considering the objective of making LIS surfaces with low clathrate hydrate adhesion, water and cyclopentane were selected as fluids. Selection of surface chemistry and texture were made such a way that lubricant under these fluids has to stabilized. Suitable LIS were fabricated by following a number of steps as mentioned below.

    **i. Selection of Lubricant:** The most important criterion in selecting the lubricant is the immiscibility of the lubricant fluid with both probe fluids i.e; water and cyclopentane. Simultaneous immiscibility with oil and water is rather challenging as liquid immiscible in water are easily miscible in oils and vice versa. Hence a suitable lubricant required to be

identified. Considering the well-known oleophobic and hydrophobic properties of fluorinated materials, a few fluorinated liquids were studied for immiscibility. To ascertain its immiscibility with water and oil, equal volume of cyclopentane (or water) and Krytox 1506 were kept in a vial and vigorously shaken for few minutes and continuously observed for one month. Within few minutes after stopping the shaking process, both phases were separated out and remain so after one month (see figure 5.a). Immiscibility and inertness of Krytox with cyclopentane hydrate is another important requirement for this specific application. These requirements are satisfied by Krytox 1506 as it showed high contact angle (>140°) on cyclopentane hydrate and no visible changes in hydrate was observed in a separate experiment. In addition to immiscibility, chemical inertness, low surface tension and low volatility are some of the other important requirements to achieve robust LIS surfaces. Krytox 1506 is a high-performance fluid which is used as lubricants and vacuum pump fluid. It has wide range of operating temperature, inertness (chemical, biological, environmental, nontoxic), significantly low surface tension compared to other liquids and very low vapor pressure desirable for LIS. Some of its important properties are listed in table 2. This krytox oil's measured interfacial tension value with cyclopentane (using Biolin Scientific Sigma 700 Force Tensiometer) was also mentioned in the table 2. Since Krytox 1506 satisfy all these important requirements of LIS lubricant, it was selected for further study.

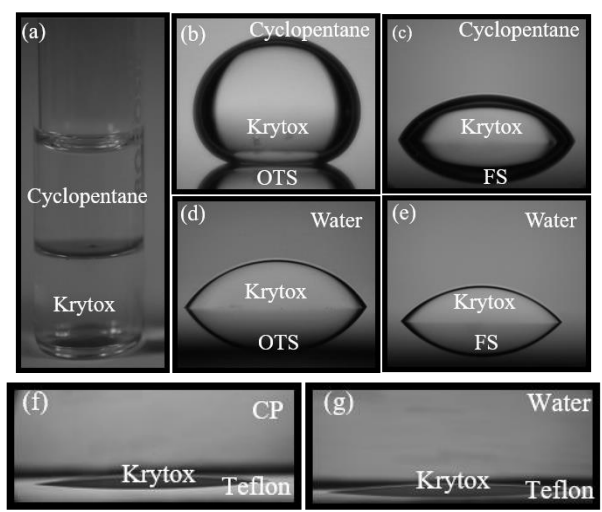

Figure 5. (a) Immiscibility of krytox with cyclopentane after one month. Contact angle of krytox in cyclopentane environment on (b) OTS surface (c) FS surface (f) Teflon surface. Contact angle of krytox in water environment on (d) OTS surface (e) FS surface (g) Teflon surface.

**ii. Selection of surface chemistry:** Silanes are frequently used to change the surface free energy of silicon/silicon dioxide and/or glass surfaces etc. Reactive sites of silanes chemically react with the surface containing OH- groups and attach its functional groups to the surfaces. Depending on the molecular structures of these functional groups surface energy modifications are achieved. In current study two different types of low surface energy silanes namely OTS, FS and teflon were used to modify the surface chemistry of the bare silicon wafer. OTS functionalization induces non polar low surface energy modifications whereas, FS functionalization induces slightly polar low surface energy modification. Based on receding contact angles, OTS and FS has free surface energy of 21 mJ/m2 and 24mJ/m2 respectively. FS expected to have higher interaction with water phase (same with hydrate) compared to nonpolar OTS due to the polar component of its surface energy. However, both surfaces showed similar water contact angles in air (105˚ ±4˚). Surface chemistry and interfacial interactions between different plays a very crucial role in design and fabrication of LIS as discussed in design and theoretical analysis section. Contact angles being the manifestation of surface chemistry and interfacial interactions, are important quantity. Here contact angles of Krytox 1506 on smooth solid surfaces in water and cyclopentane environments are most important parameter for stability analysis. The static contact angle of krytox 1506 on a smooth OTS functionalized surfaces in a cyclopentane and water environment are $160°±5°$ and $52°±3°$, respectively (see figure 5.b, 5.d). Since the presence of texture enhances nature of surface wettability, it is obvious that krytox contact angle on textured OTS surfaces will be higher than $160°$ in cyclopentane and less than $52°$ in water environment. Hence it is impossible to achieve stable impregnated krytox layer inside a OTS functionalized textured surface in cyclopentane environment. Based on the theoretical analysis performed earlier and measured contact angle values of Krytox on OTS surface in cyclopentane, it is also evident that Krytox cannot stay inside OTS functionalized textures in cyclopentane environment. Although, stable krytox layer can be obtained in same surface in water environment, absence of its stability in cyclopentane makes OTS functionalization unsuitable for omniphobic LIS fabrication. On the other hand, equilibrium contact angles of Krytox 1506 on smooth FS functionalized surfaces in both cyclopentane (72°) and water (48°) environment are less than 90° (see figure 5.c, 5.e). It implies that LIS based on FS functionalization and appropriate texture will have thermodynamically stable lubricant (Krytox) layer both in water and cyclopentane environment. However, such stable LIS surface on textured fluorosilane surface is possible with maximum post spacing of 5 micron among all fabricated micro post surfaces. Anticipating considerable hydrate adhesion forces on such surfaces due to higher area fraction of post tops, another surface chemistry modification based on liquid Teflon was employed. This coating showed significantly reduced krytox contact angles of 10° and 12° in water and cyclopentane environment respectively. Such small contact angles not only expected to give stable krytox (in water and oil) layer but also facilitate stability with higher post spacings and low hydrate adhesion. Theoretical calculations mentioned earlier predict this limit to be 100 microns. Such higher value of post spacing should reduce the hydrate adhesion force to extremely low values. Hence FS and Teflon coating were selected for further fabrication of LIS surfaces and experimental studies.

**iii. Selection of surface textures for stable LIS:** The regime map (see figure 4) showed that relationship between the smooth contact angle and the critical contact angle determines the thermodynamically stable state. For the square post geometry considered in the theoretical analysis and post dimension of 10 μm x 10 μm x 10 μm, the critical contact angle values for different post spacings were plotted (see figure 6. (a)) using a MATLAB code. The graph (see figure 6. (a)) indicates a steady declining trend, i.e., as post spacing rises, the value of $\theta_c$ decreases. By setting the measured equilibrium contact angles of krytox 1506 on FS functionalized smooth silicon surfaces (in water or cyclopentane environment) to the critical contact angle ($\theta c$, geometric property), the critical post spacing can be estimated from the same plot (see figure 6).

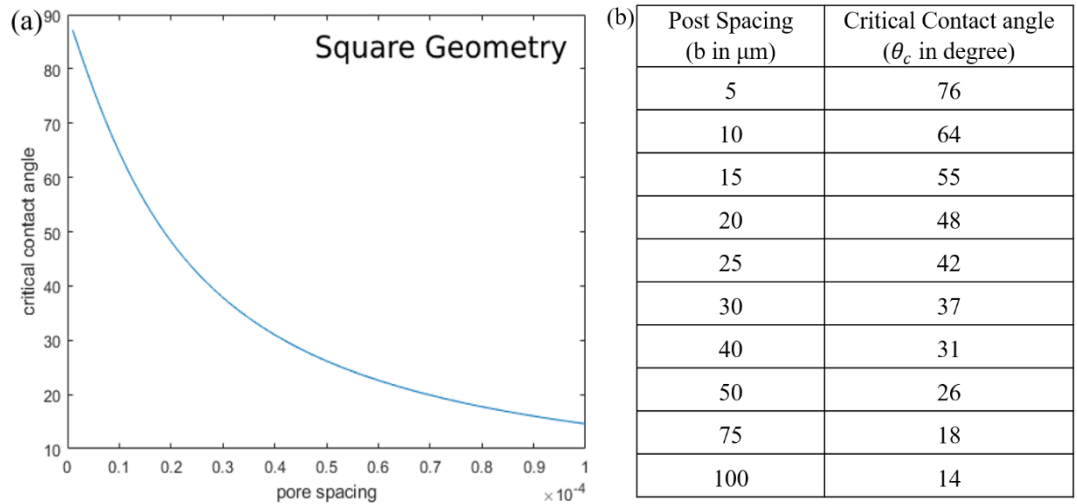

Figure 6. (a) Critical contact angle values for different post spacings were plotted for square geometry using a MATLAB code. (b) Corresponding critical contact angle values with post spacings in tabular format

Textured surfaces with post spacings less or equal to this critical post spacing expected to hold the krytox 1506 lubricant inside a given environment fluid owing to the thermodynamic stability. Surfaces with post spacing higher than this critical value will not be able to hold lubricant in its textures and lubricant expected to be replaced by the environment fluid. The exact value of critical post spacings (theoretically calculated) for textured FS surfaces was calculated to be 6.7 μm and 20 μm respectively for cyclopentane and water environment. Figure 6. (b) lists down critical contact angles of different micro post surfaces which were fabricated using optical lithography technique. A careful observation of figure 6. (b) reveals that the critical contact angles of FS micropost surfaces with 5 micron and 20 microns post spacing are just higher than $\theta_{KRY-FS(CP)}$ (72°) and $\theta_{KRY-FS(W)}$ (48°) respectively. This post spacing can be considered as critical post spacing based on available fabricated samples. It implies that Krytox 1506 will be only (among fabricated samples) stable inside the surfaces with post spacing of 5 μm under cyclopentane environment. Similarly, krytox 1506 should be stable inside the textured FS surfaces with post spacing 5, 10, 15 and 20 microns post spacings under the water environment. Since the micropost surface with 5-micron spacing expected to have thermodynamically stable lubricant layer under the water and cyclopentane environment simultaneously, it was selected to fabricate stable LIS for further experimental study. Similarly, surfaces with 50 micron post spacing was selected to fabricate LIS based on the Teflon coated textured surface (available sample with maximum post spacing). To experimentally validate the stability argument presented in this section as well as in the theoretical discussion section, micropost surfaces with of 10, 15, 20, 25, 30, 50 micron post spacings were fabricated using a standard optical lithography process. Micro posts surfaces with nano spikes post tops and 5 micron spacing were also fabricated to make LIS to facilitate reduction in hydrate surface contact area and lower adhesion. FESEM images of these surfaces are shown in figure 7.

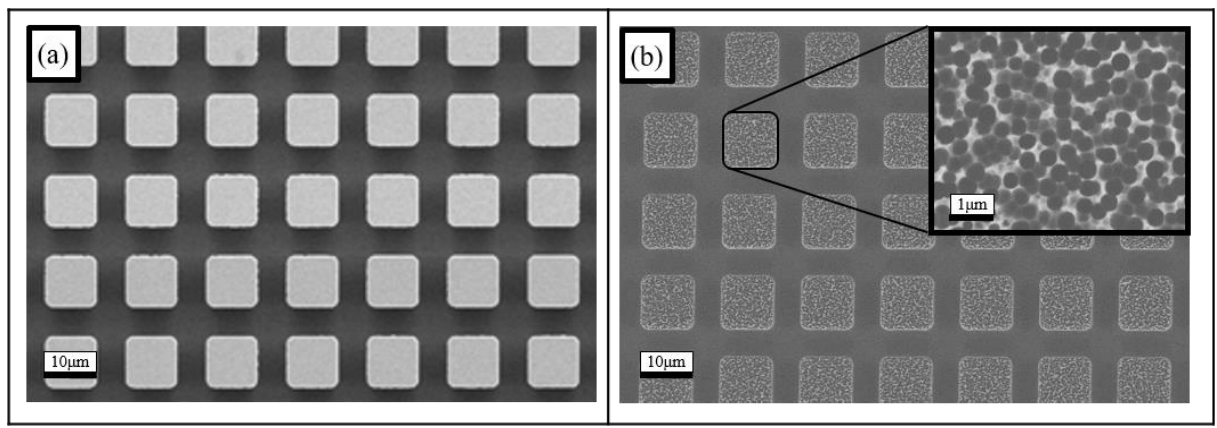

Figure 7. FESEM images (a) micropost (b) micro post with nano spike top posts

**iv. Selection of Dip Coating Speed:** Textured surfaces selected for the further studies were impregnated with the lubricant (Krytox 1506) using dip coating process. Unlike SLIPS where an arbitrarily thick layer of lubricant layer with excessive amount of lubricants is present on top of surface textures, LIS surfaces have no excess lubricant layer. To ensure the absence of excess lubricant layers, dip coating velocities were carefully selected. The capillary number values less than $10^{-4}$ corresponds to the dip coating process ensures absence of excess lubricant layer[31] irrespective of surface roughness parameters. The dip coating velocity (*V*) was obtained using the equation 7 and setting the value of capillary number to $10^{-4}$.

$$\text{Capillary number (Ca)} = \frac{\mu V}{\gamma} = 10^{-4} \tag{7}$$

Where $\mu$= viscosity of the liquid, $\gamma$= surface tension, *V*= withdrawal velocity.

The viscosity, surface tension values of the Krytox 1506 were tabulated in table 2 along the calculated withdrawal/dip coating velocity of 0.9 mm/minute. The same withdrawal velocity was used for lubricant impregnation of all textured FS functionalized surfaces to make different LIS. Absence of thick lubricating oil layer on the post tops was verified using the high zoom digital microscopic images of liquid impregnated micropost surfaces (See image S1 in Supporting Information).

Table 2. Properties and Withdrawal velocity of lubricant liquid at 20 degrees Celsius[32]

| Lubricant | Liquid-Vapour Surface Tension $(mN/m)$ | Liquid Density $(kg/m^3)$ | Dynamic Viscosity (mPa.s) | Vapour pressure $(kPa)$ | Withdrawal velocity $(mm/min)$ | Interfacial Tension $\gamma_{K-CP}$ $(mN/m)$ |
|---|---|---|---|---|---|---|
| Krytox 1506 | 17 | 1880 | 112.8 | $5 \times 10^{-8}$ | 0.9 | 3.5 |

**3.2 Experimental validation of Lubricant Stability in Micropost LIS:** Lubricant stability inside the textures of LIS is the critical requirement to maintain its functional durability for long term. Lubricant stability experiments were performed on different micro post LIS in cyclopentane and water environment to validate the theoretical analysis discussed earlier. Figure below (see figure 8) clearly shows LIS based on micropost sample with 5 micron post spacing retained lubricant inside cyclopentane environment. No visible change on this (see figure 8.a) surface observed till, the maximum experimental duration of 48 hours. In clear contrast, LIS samples made from other post spacings clearly shows the visible signs of instability as indicated by the areas inside red circles (Figure 8.b, 8.c, 8.d). Lubricant in these zones was replaced by the environmental fluid cyclopentane. Images of LIS with unstable lubricants were taken when the sign of lubricant instability were clearly visible for the first time (figure 8. b, c, d). Additional images of LIS with significant amount replaced lubricant were shown in the figure (See image S2 in Supporting Information). Onset times of this instability were found to be inversely related to the post spacing. Different time scale of lubricant replacement seems to be related to the complex dynamics of three phase contact line and outside the scope of the current study. These experimental observations accurately validate theoretical prediction on lubricant stability in LIS surfaces. Experimental study inside the water environment also showed stable lubricant layer in liquid impregnated surfaces with micropost spacings up to 20 microns for more than 48 hours. However, LIS for current application need stable lubricant layer in both cyclopentane and water. Hence, micropost LIS with 5 micron was only considered among fluorosilanized micropost LISs for further study on hydrate adhesion. Stability of LIS based on micro post (5 micron spacings) with nano spike post tops were ascertained in similar experiments and did not show any sign of instability till 48 hours. Stability of lubricant layer in these surfaces can be attributed to the presence of higher aspect ratios and hierarchical surface features evident from FESEM images. In addition to the LIS with 5-micron post spacing, these surfaces were also considered for hydrate adhesion study described in next section. Stability experiments performed on Teflon coated LIS with 50-micron post spacing showed similar long term stability in water and cyclopentane environment.

| Post Spacing | 5 µm | 10 µm | 15 µm | 20µm |
|---|---|---|---|---|
| Under Cyclopentane environment | (a) t ≥ 48 hours | (b) t = 60 mins | (c) t = 50 mins | (d) t = 20 mins |
| | **Stable** | **Unstable** | | |

Figure 8: lubricant stability/instability images under CP environment on different post spacing surfaces (a) 5µm, (b) 10µm, (c) 15µm, (d) 20µm

**3.3 Hydrate adhesion measurement:** Hydrate adhesion measurement on the selected LIS surfaces were performed in a custom made setup (see experimental section for details) using a cantilever based method. The setup's schematic was previously reported[33]. It uses the same configuration. A thin steel (SS304) wire was used to dislodge hydrate particles formed on test surfaces. Adhesion forces were calculated from the measured deflection of the wire, wire dimension and its effective elastic modulus. Except Micropost_FS_LIS, hydrate particles were dislodged from all other liquid impregnated surfaces via adhesive failure. Micropost_FS_LIS showed cohesive failure as shown in figure 9. (a) and (b). Since the adhesion force will be higher than the cohesive force for these cases, the adhesion force calculated based on the displacement will be lower than the actual adhesion. Figure 9. (c) and (d) show adhesive and cohesive failure on LIS with fluorosilanized micropost-nanospike surface. A thin layer of krytox 1506 was observed in between hydrate particles and these LIS surfaces, indicating presence of a krytox barrier film between the hydrate particle and solid texture underneath. Once the hydrate particles were removed, this lubricant meniscus were found to disappear in to the lubricant layer (See video S3 in Supporting Information). Similar observation was also made on LIS based on Teflon coating. Plot of adhesion forces on different LIS surfaces were shown in figure 9.e.

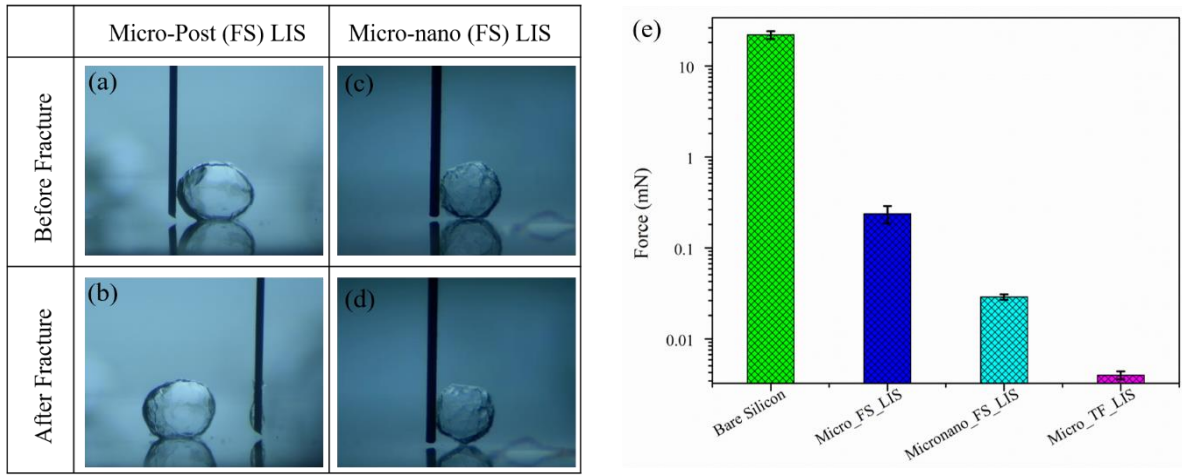

Figure 9: (a) and (b) shows both cohesive and adhesive failure on micro FS surface whereas (c) and (d) show cohesive failure on micro-nano FS surface with lower solid fraction. (e) Adhesion plot.

The adhesion force on LIS is significantly reduced compared to the same surface without out lubricant[33]. Presence of lubricant (within the texture) with lower surface tension, higher viscosity, higher density compared to cyclopentane and water possibly inhibited the growth of hydrate into the texture. As a result, the hydrate solid contact area and hydrate adhesion was reduced significantly. Presence of wedge shaped microscopic krytox layer between hydrate particles and LIS also indicates favorable interactions between hydrate and krytox to produce higher wettability or possible spreading of krytox layer on hydrate surfaces. This effect also expected to reduce the adhesion. Although the adhesion force on Micro_FS_LIS expected to be less but not extremely low. Presence of direct contact between the water/hydrate and closely spaced post tops increases overall contact area and thus results in significant adhesion force. Plotted adhesion values for these samples are underestimating the actual adhesion values. Presence of nano scale high aspect ratio spikes on Micronano_FS_LIS and higher post spacing on Micro_TF_LIS reduce the hydrate solid contact significantly leading to further reduction of the adhesion values by order of magnitudes. Hydrate particles detaches from these surfaces with adhesive failure only. The hydrate adhesion on Micro_TF_LIS with Teflon coating was exceptionally low and below the lower limit (0.004 mN) of current force measurement method. The hydrate adhesion on Micronano_FS_LIS was higher compared to the same on Micro_TF_LIS but still significantly low compared to Micro_FS_LIS . It clearly shows that tweaking the surface chemistry could reduce hydrate adhesion force by three orders of magnitude even without introducing the nano textures which may have durability issues.

**4. Conclusions:**

In conclusion, a theoretical frame work to design LIS with stable lubricant under water-oil environment and low hydrate adhesion was proposed and experimentally validated. Theoretical analysis showed the general guidelines to choose appropriate combination of surface chemistry, fluid surface energies and surface texture to achieve stable lubricant layer under different fluid or liquid environments. Equilibrium contact angle of lubricant on smooth surfaces (of given surface chemistry) under given fluid environment and critical contact angle, a purely function of surface roughness are two key parameters in determining the lubricant stability. Stability experiments performed on several LIS showed good agreement with the theoretical analysis. Hydrate adhesion measurements on several stable LIS showed reduction from one to three order of magnitude through manipulation of texture or surface chemistry. Micro_FS_LIS with micro only features though retained lubricant layer, cannot reduce the hydrate adhesion significantly due high soild fraction or solid water contact area. Cantilever used for adhesion measurement could not dislodge hydrate particles from these LIS and cohesive fracture were observed. Adhesion value reported for micro LIS was based on this failure and underestimates the actual hydrate adhesion on micro-LIS. Limitation of micro LIS was overcome by introducing high aspect ratio and low solid fraction nano-spike surface textures resulting low hydrate adhesion in the range of 0.01 mN. Micro_TF_LIS showed largest reduction due to the surface chemistry below the lower limit of measurement of 0.004 mN. Hydrate particles on Micronano_FS_LIS and Micro_TF_LIS found to be connected with the surface through a capillary bridge formed by the lubricant oil, signifying no hydrate to solid contact. Although, extremely low hydrate adhesion on LIS with nano features were observed, for practical purpose these nano features expected to be non-durable. Hence LIS with micro features and high post spacing (or low solid fraction) like Micro_TF_LIS is thus desirable for practical use as low hydrate adhesion surface. Moreover, LIS made from optical lithography as reported here is both non scalable and very expensive. To reduce the cost, LIS based on suitable low cost scalable fabrication method and low cost industrial materials are required. Currently, research activities towards such alternative surface chemistry and low cost fabrication methods are underway in our lab. Adhesion forces reported in this current report were between cyclopentane hydrate and LIS at atmospheric pressure. Although results were very encouraging, adhesion reduction of gas hydrates which forms at high pressure and low temperature condition is still required for field applications of LIS. Considering the difference in phases, surface energy parameters and their higher sensitivity with pressure and temperature conditions, a detailed study of gas hydrate adhesion on LIS under such conditions are must. For the hydrate growth or adhesion under dynamic situation can be significantly different than static condition. A custom made set up was used to evaluate hydrate accumulation on stable Micro_FS_LIS (5-micron post spacing) and control smooth silicon wafer. Hydrate formation occurred on bare silicon surfaces while barely observable on LIS surfaces. It suggests that LIS can be used to prevent hydrate plugging inside the pipe flow conditions and flow conditions will reduce the hydrate adhesion on LIS surfaces. Investigations are still continuing; however further research is required before anything substantial can be concluded about these physical phenomena.

**5. Experimental Methods:**

**5.1 Sample Cutting:** Test samples were cut from smooth or textured silicon wafers of 6-inch diameter. Wafers were cut into 20 mm by 20 mm square pieces using a 1064 nm Nd:YAG solid state laser system (Electrox). Appropriate drawings were made in a software provided with the laser system to cut wafers into maximum possible number of samples with above mentioned dimension.

**5.2 Surface Chemical Functionalization:** Solid surfaces with and without textures were chemically modified using well known silane chemistry. Two different types of low surface energy silanes, OTS (octadecyltrichlorosilane, Sigma Aldrich) or FS (tridecafluoro-1,1,2,2 tetrahydrooctyl-trichlorosilane, Sigma-Aldrich) were used for this purpose. For OTS functionalization, square cut silicon samples were first plasma cleaned for 10 mins in a RF plasma cleaner (Harrick Plasma, Model PDC-002-HP) under oxygen environment at pressure 200 m Torr. Subsequently samples were placed inside toluene (ACS grade 99%-Sigma Aldrich) containing 0.5 wt% OTS. An emulsion of 1wt% DI water (Millipore, 18.2MΩ) in toluene was prepared separately and added with OTS-Toluene solution in equal volume. The water in toluene emulsion was prepared using a high energy probe sonication (Sonics 750W sonicator) for 90s at 70% power followed by a 10 mins of sonication in a bath sonicator (Branson). Samples were kept inside the mixture of OTS, Toluene and DI water for 20 minutes to complete the reaction. Upon completion of the chemical reaction samples were bath sonicated for 3 minutes each in Acetone (ACS grade, 99% -Sigma Aldrich) and Isopropyl Alcohol (ACS grade ,99%-Sigma Aldrich) and finally rinsed with copious amount of DI water. This cleaning steps were performed to remove physically abrobed silanes on sample surfaces. Fluorosilanization process was performed using a vacuum based chemical vapor deposition technique. Samples were first treated in a plasma cleaner for 10 mins and kept inside a desiccator with 5 μL of fluorosilane at 300 mbar for 4hours. Samples were cleaned using the same protocol as mentioned earlier for the OTS functionalized samples. 1 wt% solution of Teflon AF 2400 (chemours) was used to coat the micro post surfaces followed by curing the same. Teflon AF is a high molecular weight polymer, and unless the polymer is heated above its glass transition temperature (Tg), complete solvent removal may take a long period. Coated samples were heated for 5 minutes at a temperature 5 °C above the Tg of the polymer (245 °C) to completely evaporate all solvent. Further heat the coated substrate to 330 °C for 10 to 15 minutes to improve adherence and attain the best uniformity of coating thickness.

**5.3 Surface Texturing:** Standard optical lithography was used to fabricate microposts containing surfaces on 6inch silicon wafer (n-type 1 0 0 plane, 650μm thick). Square microposts with 10μm x 10μm x 10μm dimension were fabricated with different inter post spacings between 5μm to 100μm. Shipley S1818 photoresist was used to coat the silicon wafer and coated wafer was exposed to ultraviolet light of 405nm wavelength through a chrome mask (Advance Reproductions Corporations). Exposed wafers were developed in 1:1 volume ratio of DI water and Microdev solution (Dow Chemicals). Subsequently wafer with developed photoresist layer were etched to 10 μm using an inductively coupled plasma etching system (ICPES, Surface Technology Systems). Surface Profilometry was performed using a non-contact optical profiler (CCI HD Optical Profiler, Taylor Hobson) to confirm the etched height. Finally etched wafers were cleaned in piranha solution (1:3 volume ratio of hydrogen peroxide and sulfuric acid) to remove remaining photoresists. To generate Nanograss sample thoroughly cleaned silicon wafers were directly etched in the ICPES using fluorinated gas etchants. Similarly, micropost samples were etched in ICPES using the nano grass receipe to produce micropost surfaces with nanograss tops.

**5.4 Microscopic Visualization:** The fabricated textured surfaces were characterized using a field emission scanning electron microscope (FESEM) (Sigma 300- Carl-Zeiss). A typical image was captured at around 5kV voltage and with a magnification of about 1500 x magnification. To capture nanoscale features of test samples images were taken at magnification of 15000x.

**5.5 Contact Angles Measurement:** Equilibrium, receding and advancing contact angles on various surfaces in air, water and cyclopentane environments were measured using a contact angle goniometer (Rame-Hart model 500-U1). To measure contact angles in liquid environment, test samples were submerged in a transparent quartz cell filled with environment liquid. Size of probe liquids were fixed below their respective capillary lengths (~4 μL). Dispensing and retraction rate of 0.2 μL/s was used to measure advancing and receding angles respectively. To calculate the average and standard deviation of measured quantities, at least 10 measurements from different locations of the samples were performed.

**5.6 Hydrate Accumulation:** Hydrate accumulation study was performed inside a transparent quartz cell. A propeller based mechanical mixer was used to create rotational flow of cyclopentane, water droplets inside the cuvette. Test samples were kept inside the cuvette filled with cyclopentane. Cuvette was placed on top of a Peltier cooler (CP60-TE Technology) and cyclopentane inside it was cooled to -5°C and kept for 15 mins. A 36 guage needle was used to introduce water droplets of 0.1 mm diameter or less into cold cyclopentane while it was being rotated at 200 rpm. Water flow rate was kept at 1 mL /min. Temperature of cyclopentane cuvette was cycled between 5 °C and -5 °C with ramp rate of 1 °C/ min for 30 minutes to initiate the nucleation of hydrate at the interface of ice and cyclopentane. Cyclopentane temperature was then kept at 5 °C for 90 minutes to make sure presence of only cyclopentane hydrate and absence of ice in solid phase.

**5.7 Hydrate Adhesion Force Measurement:** Hydrate adhesion measurements were carried out using a custom experimental setup. Initially a quartz cuvette functionalized with OTS was filled with cyclopentane and kept over the peltier cooler to cool down to -15°C. Test samples and few 3 μL water drops were placed inside the cyclopentane bath. Once water drops inside the cyclopentane were frozen, temperature of bath was raised to 1 °C very carefully and slowly around 0°C to ensure hydrate nucleation at water cyclopentane interface. Once hydrate covers the entire water drops, tiny water droplets of 0.12 microliter volume were deposited all over the test samples. A thin wire (0.1 mm dia) mounted on a syringe-needle system was used to collect hydrate crystal from preformed hydrate formed over larger water drops and gently introduce those hydrate crystals into the tiny water drops deposited over test samples. After introducing the hydrate crystal on such tiny water drops, system was kept at 1°C for 4 h prior to conducting adhesion force measurements. Adhesion force between the test samples and hydrate particles generated from tiny water droplets were measured from the deflection a thin SS304 stainless steel wire (Hamilton co diameter 0.1 mm). Once hydrate fully formed on the surface of tiny water droplets, the bottom end of the thin cantilever was brought in contact with hydrate particles. The cantilever was attached to a manual micro manipulator with XYZ motion and moved gently to exert force on hydrate particle until the hydrate particle detached from the test samples or fracture cohesively. A DSLR camera (Nikon D850) with a high zoom microscopic lens (Navitar 6X with 2X adapter lens) were used to record videos of the hydrate detachment from the surface with cantilever. Image analysis of these videos were performed in adobe photoshop to measure the cantilever deflection. The adhesion force was calculated from the measured deflection cantilever's material property and it's geometry using classical cantilever deflection equations.

**ASSOCIATED CONTENT**

**Supporting Information**.

The following files are available free of charge

Image S1: Absence of thick lubricating oil layer on the post tops of liquid impregnated micropost surfaces.

Image S2: Observed instability on krytox 1506 impregnated different micron surfaces under CP environment.

Video S3: Indicating presence of an krytox barrier film between the hydrate particle and solid texture. In addition, showed practically non detectable adhesion force.

## AUTHOR INFORMATION

**Corresponding Author**

* Email: arindam@iitgoa.ac.in

**AUTHOR CONTRIBUTION**

All authors contributed to the writing of the manuscript. The final document of the article has received the unanimous approval of all writers. Abhishek Mund performed theoretical calculations and data analysis. In addition, all the performed tests were carried out, designed, and the manuscript was written by Abhishek Mund. The adhesion measurement setup was made by Amit K. Nayse, who also contributed in writing the manuscript and visual design. Abhishek Mund and Amit K. Nayse both performed the adhesion measurement. Dr. Arindam Das, who also wrote the manuscript, developed the experimental design and concept.

**Funding Sources**

This research work was funded by Science Engineering Research Board (SERB) with sanction order no: SRG/2019/002011 dated: 19/11/19.

**Notes**

The authors declare no competing financial interest.

**ACKNOWLEDGMENT**

The authors thanks CoE-PCI (Centre of Excellence- Particulates, Colloids and Interface) of IIT Goa for providing the instrument facility and space to perform the experiment.

The authors thanks CNS - Center for Nanoscale Systems of Harvard University for using their cleanroom facilities to prepare lithographic samples.